# Multi-label classification for biomedical literature: an overview of the BioCreative VII LitCovid Track for COVID-19 literature topic annotations

Qingyu Chen[1,†], Alexis Allot[1,†], Robert Leaman[1], Rezarta Islamaj[1], Jingcheng Du[2], Li Fang[3], Kai Wang[3,4], Shuo Xu[5], Yuefu Zhang[5], Parsa Bagherzadeh[6], Sabine Bergler[6], Aakash Bhatnagar[7], Nidhir Bhavsar[7], Yung-Chun Chang[8], Sheng-Jie Lin[8], Wentai Tang[9], Hongtong Zhang[9], Ilija Tavchioski[10,11], Senja Pollak[11], Shubo Tian[12], Jinfeng Zhang[12], Yulia Otmakhova[13], Antonio Jimeno Yepes[14], Hang Dong[15], Honghan Wu[16], Richard Dufour[17], Yanis Labrak[18], Niladri Chatterjee[19], Kushagri Tandon[19], Fréjus A. A. Laleye[20], Loïc Rakotoson[20], Emmanuele Chersoni[21], Jinghang Gu[21], Annemarie Friedrich[23], Subhash Chandra Pujari[22,23], Mariia Chizhikova[24], Naveen Sivadasan[25], Saipradeep VG[25], Zhiyong Lu[1,*]

1. National Center for Biotechnology Information, National Library of Medicine, National Institutes of Health, Maryland, USA
2. School of Biomedical Informatics, UT Health, Texas, USA
3. Raymond G. Perelman Center for Cellular and Molecular Therapeutics, Children's Hospital of Philadelphia, Philadelphia, USA
4. Department of Pathology and Laboratory Medicine, University of Pennsylvania Perelman School of Medicine, Philadelphia, USA
5. College of Economics and Management, Beijing University of Technology, Beijing, China
6. CLaC Labs, Concordia University, Montreal, Canada
7. Navrachana University, Vadodara, India
8. Graduate Institute of Data Science, Taipei Medical University, Taipei, Taiwan
9. College of Computer Science and Technology, Dalian University of Technology, Dalian, China
10. Computer and Information Science University of Ljubljana, Ljubljana, Slovenia
11. Jožef Stefan Institute, Ljubljana, Slovenia
12. Department of Statistics, Florida State University, Tallahassee, USA
13. School of Computing and Information Systems, University of Melbourne, Melbourne, Australia
14. School of Computing Technologies, RMIT University, Melbourne, Australia
15. Centre for Medical Informatics, Usher Institute, University of Edinburgh, Edinburgh, United Kingdom
16. Institute of Health Informatics, University College London, London, United Kingdom
17. LS2N, Nantes University, Nantes, France
18. LIA, Avignon University, Avignon, France
19. Department of Mathematics, Indian Institute of Technology Delhi, New Delhi, India
20. Opscidia, Paris, France
21. Department of Chinese and Bilingual Studies, The Hong Kong Polytechnic University, Hong Kong, China
22. Heidelberg University, Heidelberg, Germany
23. Bosch Center for Artificial Intelligence, Renningen, Germany
24. SINAI Group, Department of Computer Science, Advanced Studies Center in ICT (CEATIC), Universidad de Jaén, Jaén, Spain
25. TCS Research, Life Sciences, Hyderabad, INDIA

[†] Equal contributions
* Corresponding author

## Abstract

The COVID-19 pandemic has been severely impacting global society since December 2019. The related findings such as vaccine and drug development have been reported in biomedical literature – at a rate of about 10,000 articles on COVID-19 per month. Such rapid growth significantly challenges manual curation and interpretation. For instance, LitCovid is a literature database of COVID-19-related articles in PubMed, which has accumulated more than 200,000 articles with millions of accesses each month by users worldwide. One primary curation task is to assign up to eight topics (e.g., Diagnosis and Treatment) to the articles in LitCovid. The annotated topics have been widely used for navigating the COVID literature, rapidly locating articles of interest, and other downstream studies. However, annotating the topics has been the bottleneck of manual curation. Despite the continuing advances in biomedical text mining methods, few have been dedicated to topic annotations in COVID-19 literature. To close the gap, we organized the BioCreative LitCovid track to call for a community effort to tackle automated topic annotation for COVID-19 literature. The BioCreative LitCovid dataset – consisting of over 30,000 articles with manually reviewed topics – was created for training and testing. It is one of the largest multi-label classification datasets in biomedical scientific literature. Nineteen teams worldwide participated and made 80 submissions in total. Most teams used hybrid systems based on transformers. The highest performing submissions achieved 0.8875, 0.9181, and 0.9394 for macro F1-score, micro F1-score, and instance-based F1-score, respectively. Notably, these scores are substantially higher (e.g., 12%, higher for macro F1-score) than the corresponding scores of the state-of-art multi-label classification method. The level of participation and results demonstrate a successful track and help close the gap between dataset curation and method development. The dataset is publicly available via https://ftp.ncbi.nlm.nih.gov/pub/lu/LitCovid/biocreative/ for benchmarking and further development.

## Introduction

The rapid growth of biomedical literature poses a significant challenge for manual curation and interpretation [1-3]. This challenge has become more evident during the COVID-19 pandemic: the number of COVID-19-related articles in the literature is growing by about 10,000 articles per month; the median number of new articles per day since May 2020 is 319, with a peak of over 2,500; and this volume accounts for over 7% of all of PubMed articles [4].

In response, LitCovid [5, 6], the first-of-its-kind COVID-19-specific literature resource, has been developed for tracking and curating COVID-19 related literature. Every day, it triages COVID-19-related articles from PubMed, categorizes the articles into research topics (e.g., prevention measures), and recognizes and standardizes the entities (e.g., vaccines and drugs) mentioned in each article. The collected articles and curated data in LitCovid are freely available. Since its release, LitCovid has been widely used with millions of accesses each month by users worldwide for various information needs, such as evidence attribution, drug discovery, and machine learning [6].

Initially, data curation in LitCovid was done manually with little machine assistance. The rapid growth of the COVID-19 literature significantly increased the burden of manual curation, especially for topic annotations [6]. Topic annotation in LitCovid is a standard multi-label classification task that assigns one or more labels to each article. A set of eight topics are selected for annotation based on topic modeling and discussions with physicians aiming to understand COVID-19, such as the Transmission topic, which describes the characteristics and modes of COVID-19 transmissions. The annotated topics have been demonstrated to be effective for information retrieval and have been widely used in many downstream applications. Topic-related searching and browsing account for ~20% of LitCovid user behaviors, making it the second most-used feature in LitCovid [6]. The topics have also been used in downstream studies such as citation analysis and knowledge network generation [7-9]. Figure 1 shows the characteristics of topic annotations in LitCovid.

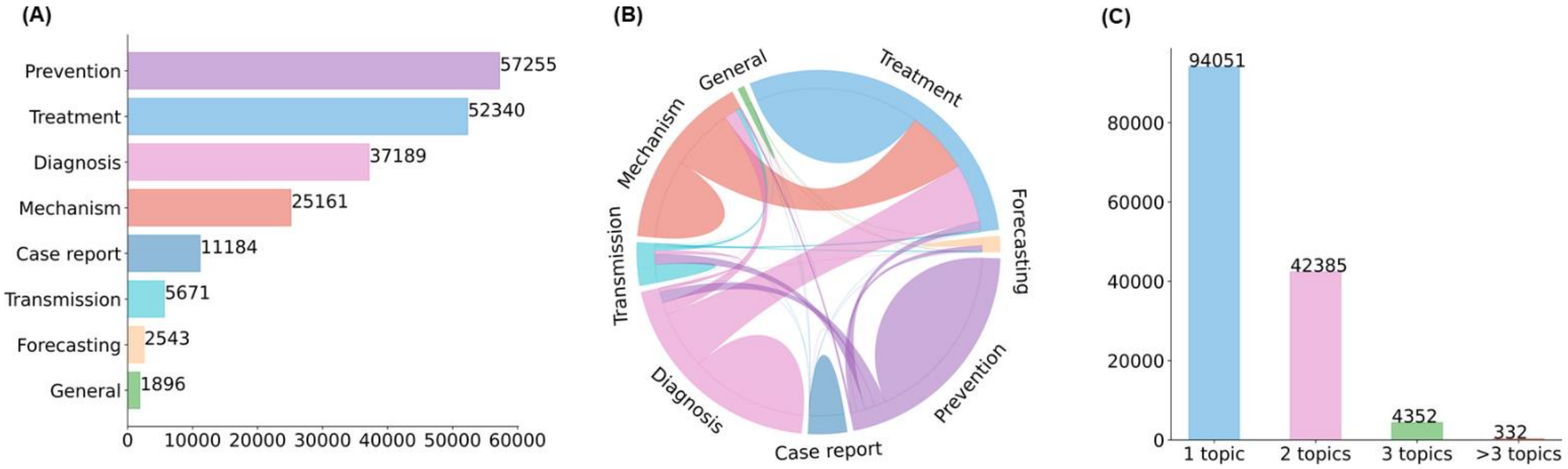

**Figure 1. Characteristics of topic annotations in LitCovid up to Feb 2022. Figure 1(A) shows the frequencies of topics; Figure 1(B) demonstrates topic co-occurrences; and Figure 1(C) illustrates the distributions of the number of topics assigned per document.**

However, annotating topics in LitCovid has been a primary bottleneck for manual curation. Compared to other curation tasks in LitCovid (document triage and entity recognition), topic annotation is more difficult due to requiring interpretation of the biomedical literature and assigning up to eight topics. As an example of the language variation that must be addressed, we provide the following five sentence snippets reflecting the treatment topic: (1) '…as a management option for COVID-19-associated diarrhea…' (PMID34741071), (2) '…modulating these factors may impact in guiding the success of vaccines and clinical outcomes in COVID-19 infections…' (PMID34738147), (3) '…lung ultrasound abnormalities are prevalent in patients with severe disease, RV involvement seems to be predictive of outcomes…' (PMID34737535), (4) '…common and virus-specific host responses and vRNA-associated proteins that variously promote or restrict viral infection…' (PMID34737357), and (5) '…the unique ATP-binding pockets on NTD/CTD may offer promising targets for design of specific anti-SARS-CoV-2 molecules to fight the pandemic…' (PMID34734665). Although these sentence snippets all describe treatment-related information, they use rather different vocabularies and structures. While automatic approaches have been developed to assist manual curation in LitCovid, the evaluations show that the automatic topic annotation tool has an F1-score of 10% lower than the tools assisting other curation tasks in LitCovid [6]. Increasing the accuracy of automated topic prediction in COVID-19-related literature would be a timely improvement beneficial to curators, biomedical researchers, and healthcare professionals.

To this end, we organized the BioCreative LitCovid track to call for a community effort to tackle automated topic annotation for COVID-19 literature. BioCreative, established in 2003, is the first and longest-running community-wide effort for assessing biomedical text mining methods [10]. Previous BioCreative challenges have successfully organized tracks on a range of biomedical text mining applications such as relation extraction [11] and entity normalization [12].

This article provides an extended overview from [13] on the BioCreative LitCovid track. It substantially describes (1) the dataset annotation characteristics, (2) detailed methods from the participating teams, and (3) in-depth evaluation results. Overall, 19 teams submitted 80 runs, and ~75% of the submissions had better performance than the baseline method [14]. The dataset and evaluation scripts are available via https://ftp.ncbi.nlm.nih.gov/pub/lu/LitCovid/biocreative/ and https://github.com/ncbi/biocreative_litcovid, respectively. We encourage further work to develop multi-label classification methods for biomedical literature.

## Dataset, baselines, and evaluation measures

### The overall LitCovid curation pipeline

The LitCovid curation pipeline has three primary modules: (1) document triage, identifying COVID-19 related articles from new articles in PubMed, (2) topic classification, assigning up to eight topics to the COVID-19 related articles (i.e., a multi-label classification task), and (3) entity recognition, extracting chemicals and locations mentioned in these articles. Initially, the curation was done manually with little machine assistance by two (part-time) human curators with a background in biomedical data sciences. As the outbreak evolved, we developed automated approaches to support manual curation and maximize curation productivity to keep up with the rapid literature growth. The detailed implementation and evaluation of the automated approaches are fully described in the description of the LitCovid resource [6]. In summary, all automated methods were evaluated before first use and have been improved continuously. The evaluations demonstrated that automated methods can achieve exceptionally high performance for document triage and entity recognition (e.g., the F1 scores were 0.99 and 0.94 for document triage and entity recognition, respectively). In contrast, the F1 score of the topic classification was 0.80, largely due to the complexity of the multi-label classification task, which assigns up to eight topics. We therefore organized this to call for a community effort to tackle automated topic annotation for COVID-19 literature.

### Topic annotations in LitCovid

The topic annotation step assigns up to eight topics to the COVID-19 related articles:

1. <u>Case Report</u>: descriptions of specific patient cases related to COVID-19,
2. <u>Diagnosis</u>: COVID-19 assessment through symptoms, test results, and radiological features for COVID-19,
3. <u>Epidemic Forecasting</u>: estimation on the trend of COVID-19 spread and related modeling approach,
4. <u>General Information</u>: COVID-19 related brief reports and news,
5. <u>Mechanism</u>: underlying cause(s) of covid-19 infections and transmission and possible drug mechanism of action,
6. <u>Prevention</u>: prevention, control, mitigation, and management strategies,
7. <u>Transmission</u>: characteristics and modes of COVID-19 transmissions,
8. <u>Treatment</u>: treatment strategies, therapeutic procedures, and vaccine development for COVID-19.

Note that by design Case Report and General Information are singleton topics, i.e., not co-assigned with other topics. This is due to their broad scope, e.g., a case report typically also contains diagnostic information.

Topics are annotated mainly based on titles and abstracts of the papers; the curators may also look for other information such as full-text and Medical Subject Headings (MeSH) when needed. Previous studies have shown that many COVID-19 articles published in PubMed without abstract information are not descriptions of formal research studies but rather commentary or perspective [15]. We also find that automatic topic annotation methods achieve 10% higher F1-score on articles with abstracts available [6]. Since late August 2020, we have prioritized annotating topics for the articles with abstracts available in PubMed, when the number of daily new articles reached a record high of over 2500.

Table 1. BioCreative LitCovid dataset characteristics in comparison with representative multi-label classification datasets on biomedical scientific literature. Note that the Chemical Exposure dataset does not provide dataset splits.

| | Dataset scale | | | | Label scale | | | Annotator scale |
|---|---|---|---|---|---|---|---|---|
| | Total documents | Train | Valid | Test | Total labels | Avg labels per doc | Unique labels | Annotators |
| Hallmarks of Cancer [16] | 1,580 | 1,108 | 157 | 315 | 2,469 | 1.56 | 10 | 1 |
| Chemical Exposure [17] | 3,661 | - | - | - | 21,233 | 5.80 | 32 | 1 |
| **BioCreative LitCovid (ours)** | 33,699 | 24,960 | 6,239 | 2,500 | 46,368 | 1.38 | 7 | 2 |

Table 2. Detailed topic annotation characteristics. Note that the General Information topic is excluded as the annotation priority is given to the articles with abstracts available in PubMed.

| | Train | | Valid | | Test | | All | |
|---|---|---|---|---|---|---|---|---|
| | #Articles | Label (%) | #Articles | Label (%) | #Articles | Label (%) | #Articles | Label (%) |
| Case Report | 2,063 | (8.27%) | 482 | (7.73%) | 197 | (7.88%) | 2,742 | (8.14%) |
| Diagnosis | 6,193 | (24.81%) | 1,546 | (24.78%) | 722 | (28.88%) | 8,461 | (25.11%) |
| Epidemic Forecasting | 645 | (2.58%) | 192 | (3.08%) | 41 | (1.64%) | 878 | (2.61%) |
| Mechanism | 4,438 | (17.78%) | 1,073 | (17.2%) | 567 | (22.68%) | 6,078 | (18.04%) |
| Prevention | 11,102 | (44.48%) | 2,750 | (44.08%) | 926 | (37.04%) | 14,778 | (43.85%) |
| Transmission | 1,088 | (4.36%) | 256 | (4.1%) | 128 | (5.12%) | 1,472 | (4.37%) |
| Treatment | 8,717 | (34.92%) | 2,207 | (35.37%) | 1,035 | (41.4%) | 11,959 | (35.49%) |

Table 3. Inter-annotator agreement on a random sample of 200 articles. Note that the General Information topic is excluded as the annotation priority is given to the articles with abstracts available in PubMed.

| Topic | Size (percentage) | Pearson correlation |
|---|---|---|
| Case report | 15 (7.50%) | 0.90 |
| Diagnosis | 37 (18.50%) | 0.71 |
| Epidemic Forecasting | 5 (2.50%) | 0.51 |
| Mechanism | 35 (17.50%) | 0.72 |
| Prevention | 94 (47.00%) | 0.84 |
| Transmission | 4 (2.00%) | 0.66 |
| Treatment | 66 (33.00%) | 0.77 |
| Macro-average | - | 0.73 |
| Micro-average | - | 0.78 |

## Dataset characteristics

Table 1 summarizes the dataset characteristics in terms of the scale of the dataset, labels, and annotators. It also compares the dataset with representative counterparts. There are only a few existing multi-label classification datasets for biomedical scientific literature, and their size is relatively small. The Hallmarks of Cancer dataset [16] has been widely used for multi-label classification methods, which has about ~1,600 documents. Another dataset on chemical exposure assessment [17] has ~3,700 documents. In contrast, The BioCreative LitCovid dataset has ~34,000 documents in total, which is nearly 10 times larger. The training, development, and testing sets contain 24,960, 6,239, and 2,500 articles in LitCovid, respectively. Table 2 shows the detailed topic distributions of the dataset. The topics were assigned using the above annotation approach consistently. All the articles contain both titles and abstracts available in PubMed and have been manually reviewed by curators. The only difference is that the datasets do not contain the General Information topic since the priority is given to the articles with abstracts available in PubMed. The training and development datasets were made available on June 15, 2021, to all participant teams. The testing set contains held-out articles added to LitCovid from June 16 to August 22, 2021. Using incoming articles to generate the testing set facilitates the evaluation of the generalization capability of automatic tools.

In addition, most existing multi-label datasets on biomedical literature were annotated by a single curator, which does not allow inter-annotator agreement to be measured. A random sample of 200 articles in LitCovid was used to measure inter-annotator agreement, and two curators annotated each article independently. Table 3 shows that the micro-average of Pearson correlation of the curators across the seven topics is 0.78, which can be interpreted as 'strong correlation' [18]. The distribution of the topics in the random sample is also consistent with that of the entire dataset. Given the scale of the dataset, the curator each annotated half of the remaining dataset and discussed difficult cases together.

### Baseline method

We chose ML-Net [14] as the baseline method. ML-Net is a deep learning framework specifically for multi-label classification tasks for biomedical literature. It has achieved favorable state of the art (SOTA) performance in a few biomedical multi-label text classification tasks and its source code is publicly available [14]. ML-Net first maps texts into high dimensional vectors through deep contextualized word representations (ELMo) [19], and then combines a label prediction network and label count prediction to infer an optimal set of labels for each document. We ran ML-Net with ten different random seeds and reported the median performance.

### Evaluation measures

Evaluation measures for multi-label classification tasks can be broadly divided into two groups: (1) label-based measures, which evaluate the classifier's performance on each label, and (2) instance-based measures (also called example-based measures), which aim to evaluate the multi-label classifier's performance on each test instance [20-22]. Both groups have unique strengths and complement each other: label-based measures quantify the effectiveness of each individual label, whereas instance-based measures quantify the effectiveness of instances which may contain multiple labels. We employed representative metrics from both groups to provide a broader evaluation of the performance. Specifically, for label-based measures, we calculated macro and micro averages on Precision, Recall, and F1-score; for instance-based measures, we calculated instance-based Precision, Recall, and F1-score. Out of these nine metrics, we focus on the three F1-scores because these aggregate both Precision and Recall.

Table 4. Team participation details, ordered alphabetically by team name.

| Team name | Team affiliation | Submissions |
|---|---|---|
| Bioformer | Children's Hospital of Philadelphia | 5 |
| BJUT-BJFU | Beijing University of Technology and Beijing Forestry University | 5 |
| CLaC | Concordia University | 4 |
| CUNI-NU | Navrachana University and Charles University | 5 |
| DonutNLP | Taipei Medical University, Taipei Medical University Hospital, and National Tsing Hua University | 5 |
| DUT914 | Dalian University of Technology | 3 |
| E8@IJS | Jozef Stefan Institute | 3 |
| ElsevierHealthSciences | Elsevier | 1 |
| FSU2021 | Florida State University | 5 |
| ittc | University of Melbourne and RMIT University | 4 |
| KnowLab | University of Edinburgh and University College London | 5 |
| LIA/LS2N | Avignon Université | 4 |
| LRL_NC | Indian Institute of Technology Delhi | 5 |
| Opscidia | Opscidia | 5 |
| PIDNA | Roche Holding Ltd | 3 |
| polyu_cbsnlp | The Hong Kong Polytechnic University and Tencent AI Lab | 5 |
| robert-nlp | Bosch Center for Artificial Intelligence and Bosch Global | 5 |
| SINAI | Universidad de Jaén | 4 |
| TCSR | Tata Consultancy Services | 4 |

Table 5. Systems and performance. The systems are categorized in terms of additional training data and knowledge sources, backbone models, and methods. The best performance in terms of each metric is also reported.

| Team name | Systems | | Best performance | | |
|---|---|---|---|---|---|
| | Additional training data and knowledge sources | Models and methods | Micro F1 | Macro F1 | Instance F1 |
| Bioformer | - | BioBERT, PubMedBERT, and Bioformer | 0.9181 | 0.8875 | 0.9334 |
| BJUT-BJFU | - | FastText, TextRCNN, TextCNN, Transformer, and correlation learning | 0.8556 | 0.7847 | 0.8701 |
| CLaC | DrugBank and MeSH | Multi-input RIM model and ClinicalBERT | 0.8897 | 0.8487 | 0.9102 |
| CUNI-NU | - | SciBERT, dual attention modules, and Label-Wise-Attention-Network | 0.8959 | 0.8673 | 0.9153 |
| DonutNLP | - | BioBERT and ensemble learning | 0.9174 | 0.8754 | 0.9346 |
| DUT914 | - | BioBERT and label feature enhancement module | 0.9175 | 0.8760 | 0.9394 |
| E8@IJS | - | Autoboot and doc2vec | 0.8430 | 0.7382 | 0.8518 |
| FSU2021 | - | PubMedBERT and multi-instance learning | 0.9067 | 0.8670 | 0.9247 |
| ITTC | - | SVM, SciBERT, Specter, BioELECTRA, and ensemble learning | 0.9000 | 0.8669 | 0.9185 |
| KnowLab | Back translation (to German), keywords, journals, UMLS, MeSH, SJR journal categories | BlueBERT-Base, PubMedBERT, JMAN, HLAN, HA-GRU, HAN, CNN, LSTM, and ensemble learning | 0.8932 | 0.8601 | 0.9169 |
| LIA/LS2N | - | TARS transformer, few-shot learning, and TF-IDF | 0.8830 | 0.8366 | 0.9094 |
| LRL_NC | - | Co-occurrence learning, TF-IDF, and LGBM | 0.8568 | 0.7742 | 0.8830 |
| Opscidia | - | BERT, data augmentation, and ensemble learning | 0.9135 | 0.8824 | 0.9296 |
| polyu_cbsnlp | MeSH | BioBERT-Base, BioBERT-Large, PubMedBERT, CovidBERT, BioELECTRA, BioM-ELECTRA, BioMed_RoBERTa, and ensemble learning | 0.9139 | 0.8749 | 0.9319 |
| robert-nlp | Publication type, keywords, and journals | SciBERT | 0.9032 | 0.8655 | 0.9251 |
| SINAI | Synonyms from WordNet | Logistic regression and TF-IDF | 0.8254 | 0.7643 | 0.8086 |
| TCSR | Biomedical entities | BioBERT and ensemble learning | 0.8495 | 0.7896 | 0.8845 |

## Results and discussions

### Participating teams

Table 4 provides details on the participating teams and their number of submissions. Each team is allowed to submit up to five test set predictions. Overall, 19 teams submitted 80 valid testing set predictions in total.

### System descriptions

Out of 19 teams, 17 teams agreed to participate in the track overview and described their approaches. Table 5 summarizes their methods and associated performance. The full detail is further provided in Table S1. Overall, we notice that the transformer approach has been used extensively: 14 out of the 17 teams (82.3%) used transformers purely (nine teams) and a combination of transformers and other traditional deep learning approaches (five teams). In contrast, only two teams used deep learning approaches besides transformers only, and two teams used machine learning approaches only. This is different from previous BioCreative challenge tasks, where most teams used machine learning approaches or a combination of machine learning and deep learning techniques [11, 23-26]. In addition, of the 14 teams using the transformer approach, seven teams (50%) proposed innovative methods beyond the default approach (fine-tuning the transformers). For instance, the Bioformer team proposed a lightweight transformer architecture which reduces the number of parameters by two thirds (the detail is summarized in [27]); the DUT914 team proposed an enhanced transformer model which learns the correlations between labels for the multi-label classification task (the detail is summarized in [28]). Such innovative approaches demonstrated superior performance and achieved top-ranked results. In addition, six teams (35%) used additional data (beyond titles and abstracts) for training the models, including metadata (e.g., paper types and journals), entity annotations (e.g., UMLS [29] and DrugBank [30]), and synonyms (e.g., WordNet [31]). The individual approaches are detailed below.

*Bioformer team* [27]

We performed topic classification using three BERT models: BioBERT [32], PubMedBERT [33], and Bioformer (https://github.com/WGLab/bioformer/). For BioBERT, we used $\text{BioBERT}_{\text{Base-v1.1}}$, which is the version described in the publication [32]. PubMedBERT has two versions: one version was pre-trained on PubMed abstracts (denoted by $\text{PubMedBERT}_{\text{Ab}}$ in this study), and the other version was pre-trained on PubMed abstracts plus PMC full texts (denoted by $\text{PubMedBERT}_{\text{AbFull}}$). We used $\text{Bioformer}_{\text{8L}}$, which is pretrained on PubMed abstracts and one million PMC full-text articles for 2 million steps. We formulated the topic classification task as a sentence pair classification problem where the title is the first sentence and the abstract is the second sentence. The input is represented as “[CLS] title [SEP] abstract [SEP].” The representation of the [CLS] token in the last layer was used to classify the relations. We utilized the sentence classifier in the transformers python library to fine-tune the models. We treated each topic independently and fine-tuned seven different models (one per topic). We fine-tuned each BERT model on the training dataset for three epochs. The maximum input sequence length was fixed to 512. We selected a batch size of 16 and a learning rate of 3e-5.

*BJUT-BJFU team* [34]

We combined the training and development sets to create our training set, which we further grouped into ten disjoint subsets with nearly equal size and similar label distribution using the stratification method in Sechidis et al. [35]. Our method takes advantage of four powerful deep learning models: FastText [36], TextRCNN [37], TextCNN [38] , and Transformer [39]. We also consider the correlations among labels [40].

*CLaC team* [41]

We used a multi-label classification approach, where a base network (shared by several classifiers) is responsible for representation learning for all classes. Although the classes might be related, different classes often require focus on different parts of the input. To allow a differential focus on the input, we used the multi-input RIM model [42] with 7 class modules, one for each class, each using ClinicalBERT [43] as input. We also used a gazetteer module for leveraging annotations from DrugBank [30] and MeSH [44]. The modules sparsely interact with one another through an attention bottleneck, enabling the system to achieve compositional behavior. The proposed model improves all classes, especially the two least frequent classes, Transmission and Epidemic forecasting. Moreover, the functionality of the modules is transparent for inspection [45].

*CUNI-NU team* [46]

Our approach implemented the SPECTER model [47], which incorporates SciBERT [48] to produce the document-level embedding using citation-based transformers. SciBERT can decipher the dense biomedical vocabulary in the COVID-19 literature, making it a valuable choice. Furthermore, we used a dual attention module [39], consisting of two self-attention layers applied to the embeddings in sequential order. These self-attention layers allow each input to establish relationships with other instances. To obtain unique vectors, i.e., query (Q), key (K), and value (V), three individually learned matrices are multiplied with the input vector. A single self-attention layer can learn the relationship between contextual semantics and sentimental tendency information. The dual self-attention mechanism helps retain more information from the sentence and thus generates a more representative feature vector. However, the dual self-attention mechanism can only generate relationships amongst the input instances while completely discarding the output. A Label-Wise-Attention-Network (LWAN) [49] is used to improve the results further and overcome the limitation of dual-attention. LWAN provides attention to each label in the dataset and improves individual word predictability by paying special attention to the output labels. It uses attention to allow the model to focus on specific words in the input rather than memorizing the essential features in a fixed-length vector. Label-wise attention mechanism repeatedly applies attention L (number of labels) times, where each attention module is reserved for a specific label. Weighted binary cross-entropy is used as a loss function. This loss function was most appropriate as it gives equal importance to the different classes during training, which was necessary due to the significant imbalance in the data. Thus, this approach overcame the significant imbalance amongst class labels and attained extensive results on labels like Case study, Epidemic forecasting, Transmission, and Diagnosis.

*DonutNLP team* [50]

We proposed a BERT-based Ensemble Learning Approach to predict topics for the COVID-19 literature. To select the best BERT model for this task, we conducted experiments estimating the performance of several BERT models using training data. The results demonstrate that BioBERTv1.2 achieved the best performance out of all models. We then used ensemble learning with a majority voting mechanism to integrate multiple BioBERT models, which are selected by the results of k-fold cross-validation. Finally, our proposed method can achieve remarkable performances on the official dataset with precision, recall, and F1-score of 0.9440, 0.9254, and 0.9346, respectively.

*DUT914 team* [28]

We designed a feature enhancement approach to address the problem of insufficient features in medical datasets. First, we extract the article titles and the abstracts from the dataset. Then the article title and the abstract are concatenated as the first input part. We only take the article titles as the second input part. Additionally, we count the distribution of labels in the training set and design a tag association matrix based on the distribution. Second, we process the features to achieve feature equalization. The first input part is tokenized and then encoded by the pretrained model BioBERT [32]. The second input part is embedded randomly. Then, we concatenate the processed features and the tokenization of the title to obtain the equalized features. Finally, we design a feature enhancement module to integrate the previously obtained label features into the model. We multiply the equalized features by the label matrix to obtain the final output vector used for classification.

*E8@IJS team* [51]

Our approach [51] used the autoBOT (automated Bag-Of-Tokens) system by Škrlj et al. [52] with some task-specific modifications. The main idea of the autoBOT system is representation evolution by learning the weights of different representations, including token, sub-word, and sentence-level (contextual and non-contextual) features. The system produces a final representation that is suitable for the specific task.

First, we transformed the multi-label classification task into a binary classification problem by treating each assignable topic as a binary classification. Next, we developed three configurations of the autoBOT system. The first configuration Neural includes two doc2vec-based latent representations, each with a dimensionality of 512. The second configuration, Neurosymbolic-0.1, includes both symbolic and sub-symbolic features, where the symbolic features include features based on words, characters, part-of-speech tags, and keywords; the dimension of the symbolic feature subspaces is 5,120. The third configuration, Neurosymbolic-0.02, has symbolic and sub-symbolic features, the same as the second configuration, but the dimensionality of the symbolic feature subspaces is 25,600.

Even if the organizers' baseline model [14] has better performance in most of the metrics, the Neurosymbolic-0.1 configuration of the autoBOT system achieves label-based micro- and macro-precision of 0.8930 and 0.9175, respectively, by which it outperforms the baseline system (for 8 percentage points in terms of macro precision). Moreover, by our results of label-based F1-score (micro) of 0.8430 (Neurosymbolic-0.02 configuration) and F1-score (macro) of 0.7382 (Neural configuration), the system has results comparable to the state-of-the-art baseline system (cca. 2% below), which indicates that autoML is a promising path for future work.

*FSU2021 team* [53]

In our participation in the BioCreative VII LitCovid track, we evaluated several deep learning models built on PubMedBERT, a pre-trained language model, with different strategies to address the challenges of the task. Specifically, we used multi-instance learning to deal with the large variation in the lengths of the articles and used the focal loss function to address the imbalance in the distribution of different topics. We also used an ensemble strategy to achieve the best performance among all the models. Test results of our submissions showed that our approach achieved a satisfactory performance with an F1 score of 0.9247, which is significantly better than the baseline model (F1 score: 0.8678) and the average of all the submissions (F1 score: 0.8931).

*ITTC team* [54]

Team ITTC combined traditional bag-of-words classifiers such as their implementation of MTI ML (a linear SVM model using gradient descent and the modified Huber loss [55, 56], available at https://github.com/READ-BioMed/MTIMLExtension), and neural models including SciBERT [48], Specter [47] and BioELECTRA [57]. We combined these into two ensemble methods: averaging across the results of SciBERT, MTI ML, and Specter, on the one hand, and taking the maximum of scores assigned by SciBERT and MT ML, on the other. The reason for such ensembles was that SciBERT tended to give high scores to well-represented categories such as Treatment while assigning scores close to zero for weaker classes such as Transmission, so its performance varied greatly depending on the composition of the test set. Conversely, Specter and MTI ML were more conservative but assigned more scores close to 0.5 even for underrepresented labels which improved precision for difficult categories. The ensemble based on the maximum value proved to be an effective strategy for recall, while averaging improved precision, especially for under-represented and challenging categories, which led to very strong macro precision results.

*KnowLab team* [58]

KnowLab group applied deep-learning based document classification models, including BlueBERT-Base [59] and PubMedBERT [33], JMAN [60], HLAN [61], HA-GRU [62], HAN [63], CNN [38], LSTM [64], etc., and each with a different combination of metadata (title, abstract, keywords, and journal name), knowledge sources (UMLS, MeSH, and SJR journal categories), pre-trained embeddings, and data augmentation with back translation (to German). A class-specific ensemble averaging of the top-5 models was then applied. The overall approach achieved micro-F1 scores of 0.9031 on the validation set and 0.8932 on the test set.

*LIA/LS2N team* [65]

We addressed the multi-label topic classification problem by combining an original keyword enhancement method with the TARS transformer-based approach [66] designed to perform few-shot learning. This model has first the advantage of not being constrained by the class number using a binary-like classification. Second, it tries to integrate the semantic information of the targeted class name in the training process by linking it to the content. Our best system architecture then uses a TARS model fed with various textual data sources such as abstracts, titles, and keywords. Then, we applied a keyword-based enhancement that consists in applying a first term frequency-inverse document frequency (TF-IDF) pass on the data to extract the specific terms of each topic with a score greater than 0.65. These terms are then framed by tags [67], the idea being to explicitly give more importance to these terms during their modeling by the TARS model. Experiments conducted during the BioCreative challenge on the multi-label classification task show that our approach outperforms the baseline (ML-Net), no matter the metric considered, while being close to the best challenge approaches.

*LRL_NC team* [68]

We propose two main techniques for this challenge task. The first technique is a data-centric approach which uses insights on label co-occurrence patterns from the training data to segment the given problem into sub-problems. The second technique uses document-topic distribution extracted from contextual topic models as features for a binary relevance multi-label classifier. The best performance across different metrics was obtained using the first technique with TF-IDF representation of the raw text corpus as features. To solve each of these multi-label classification sub-problems, Random k-Labelsets (RAKEL) classifier [69] was used with LGBM [70] as base estimator.

*Opscidia team* [71]

We propose creating an ensemble model by aggregating the sub-models at the end of each fine-tuning epoch with a weighting related to the Hamming loss. These models, based on BERT, are first pre-trained on heterogeneous corpora in the scientific domain. The resulting meta-model is fed with several semi-independent samples augmented by random masking of COVID-19 terms, the addition of noise, and the replacement of expressions with similar semantic features. While it is resource intensive if used directly, we consider its purpose to be distilling its rich new representation into a faster model.

*polyu_cbsnlp team* [72]

We propose an ensemble learning-based method that utilizes multiple biomedical pre-trained models. Specifically, we propose to ensemble seven advanced pre-trained models for the LitCovid multi-label classification problem, including BioBERT-Base [32], BioBERT-Large [32], PubMedBERT [33], CovidBERT [73], BioELECTRA [57], BioM-ELECTRA [6], and BioMed_RoBERTa [74], respectively. The homogeneous and heterogeneous neural architectures of these pretraining models assure the diversity and robustness of the proposed method. Furthermore, the extra biomedical knowledge of MeSH terms is also employed to enhance the semantic representations of the ensemble learning method. The final experimental results on the LitCovid shared task show the effectiveness and success of our proposed approach.

*robert-nlp team* [75]

Our system represents documents using n-dimensional vectors using textual content (title and abstract) and metadata fields (pubtype, keywords, and journal). Textual content and keywords are each encoded with SciBERT [48], and the two embeddings are concatenated. Following [76], this document representation is fed into a classification layer comprised of several multi-layer perceptrons, each predicting the applicability of a single label. The model outperforms the shared task baseline both in terms of macro-F1 and in terms of micro-F1. Also, it is at par with the Q3 of the task statistics, which means that results are better than 75% of all the submitted runs.

*SINAI team* [77]

To address the task of multi-label topic classification for COVID-19 literature annotation, the SINAI team opted for a problem transformation method that considers the prediction of each label as an independent binary classification task. This approach allowed the team to use the Logistic Regression algorithm [78] based on TF-IDF [79] representation of the tokenized and stemmed text data, which was previously subjected to a corpus augmentation process. This process consisted of using such techniques as back-translation [80] of a selection of articles tagged with the less represented labels (Transmission, Case Report, and Epidemic Forecasting) and the replacement of all nouns present in the abstracts with their synonyms retrieved from the WordNet [31]. The classifier achieved 0.91 label-based micro average precision and negligible time and computational resources required to train our classifier addresses the fast growth of LitCovid.

*TCS Research team* [81]

We propose two different approaches for the task. The first approach, System 1, uses the training and validation datasets directly, whereas the second approach, System 2, performs named entity recognition (NER) on the training and validation datasets and uses the resulting tagged data for training/validation. NER on the abstract and title texts was performed using our text-mining framework PRIORI-T [82], where we cover 27 different entity types, including human genes, SARS/MERS/SARS-CoV-2 genes, phenotypes, drugs, diseases, GO terms, etc. In both approaches, training is performed by fine-tuning a BioBERT model pretrained on the MNLI corpus [83]. Two separate BioBERT [32] fine-tuned models were created; the first model uses only the 'abstract' part of the training data, the second model uses only the 'remaining' part of the text, consisting of article title and metadata such as keywords and journal type. The final prediction was obtained by combining the predictions of both models, meaning that System 1 and System 2 each consist of a separate ensemble model. System 1 showed better performance than System 2 on both label and instance based F1 scores. Furthermore, System 1 showed better label-based macro and instance-based F1 scores than the challenge baseline model (ML-Net) [14]. Finally, as per the challenge benchmarks, the label-based macro F1-score for System 1 was close to the median F1 score and the instance-based F1-score was close to the mean score.

Table 5. Overall team submission-related statistics and the baseline performance. The baseline performance is the median of ten repetitions using different random seeds.

| | Label-based | | Instance-based |
|---|---|---|---|
| | Macro F1 | Micro F1 | F1 |
| **Teams** | | | |
| Mean | 0.8191 | 0.8778 | 0.8931 |
| Q1 | 0.7651 | 0.8541 | 0.8668 |
| Median | 0.8527 | 0.8925 | 0.9132 |
| Q3 | 0.8670 | 0.9083 | 0.9254 |
| **Baseline** | | | |
| ML-Net | 0.7655 | 0.8437 | 0.8678 |

Table 6. Top 5 team submission results ranked by each F1-score measure.

| Label-based | | | | Instance-based | |
|---|---|---|---|---|---|
| Macro F1 | | Micro F1 | | F1 | |
| Team | Result | Team | Result | Team | Result |
| Bioformer | 0.8875 | Bioformer | 0.9181 | DUT914 | 0.9394 |
| Opscidia | 0.8824 | DUT914 | 0.9175 | DonutNLP | 0.9346 |
| DUT914 | 0.8760 | DonutNLP | 0.9174 | Bioformer | 0.9334 |
| DonutNLP | 0.8754 | polyu_cbsnlp | 0.9139 | polyu_cbsnlp | 0.9321 |
| polyu_cbsnlp | 0.8749 | Opscidia | 0.9135 | ElsevierHealth Sciences | 0.9307 |

## Evaluation results

Table 6 summarizes team submission-related statistics and the baseline performance in terms of their macro F1-score, micro F1-score, and instance-based F1-score. The detailed results for each team submission and all the measures are provided in Table S1 in the supplementary material. The average macro F1-score, micro F1-scores, and instance-based F1-scores are 0.8191, 0.8778, and 0.8931, respectively, all higher than the respective baseline scores. The baseline performance is close to the Q1 statistics for all the three measures, suggesting that ~75% of the team submissions have better performance than the baseline method.

Figure 2 and Figure 3 further show the distributions of the overall performance and individual topic performance, respectively. Out of the seven topics, the teams achieved higher performance in terms of the median F1-score in six topics than the baseline (up to 29% higher) except the Prevention topic (only 4% lower). The results show that the performance difference is larger in the topics with relatively lower frequencies: Epidemic Forecasting (23% higher) and Transmission (29% higher). In addition, we observe that the teams achieved generally consistent performance with the correlation of manual annotations in Table 3. For instance, it had the lowest performance on the Transmission topic, which is consistent with the correlation of manual annotations in Table 3. The only exception is the Epidemic Forecasting topic, where the inter-annotator agreement had a correlation of over 0.5, whereas the teams achieved an F1-score of over 0.9. This is primarily because of the sample size: only five and 41 articles are annotated with the Epidemic Forecasting topic in the random sample for inter-annotation agreement and the entire testing set, respectively. Given the limited size, we believe the performance on the Epidemic Forecasting topic is less representative. In contrast, other topics (which have a higher number of instances) show consistent performance.

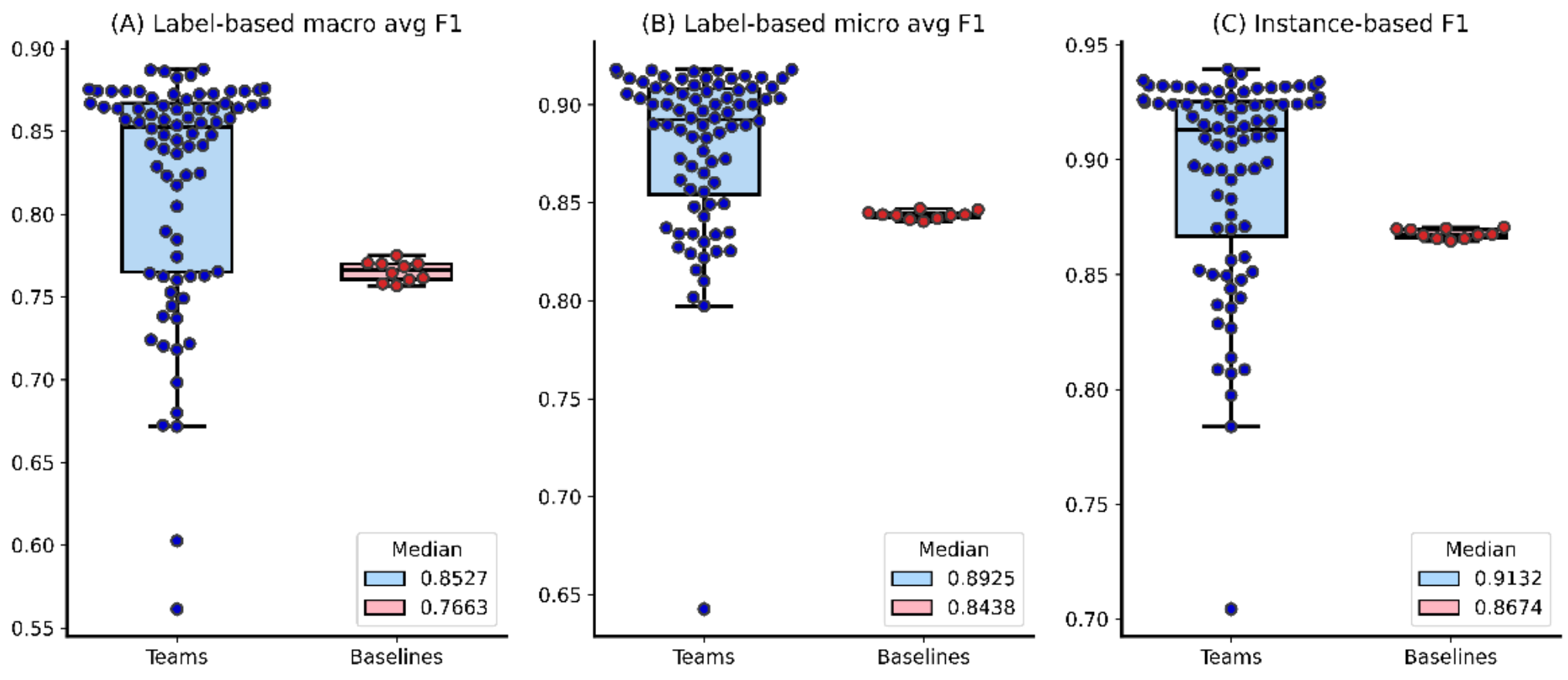

Figure 2. The distributions of team submission and baseline F1-scores. Median F1-scores are shown in the legend.

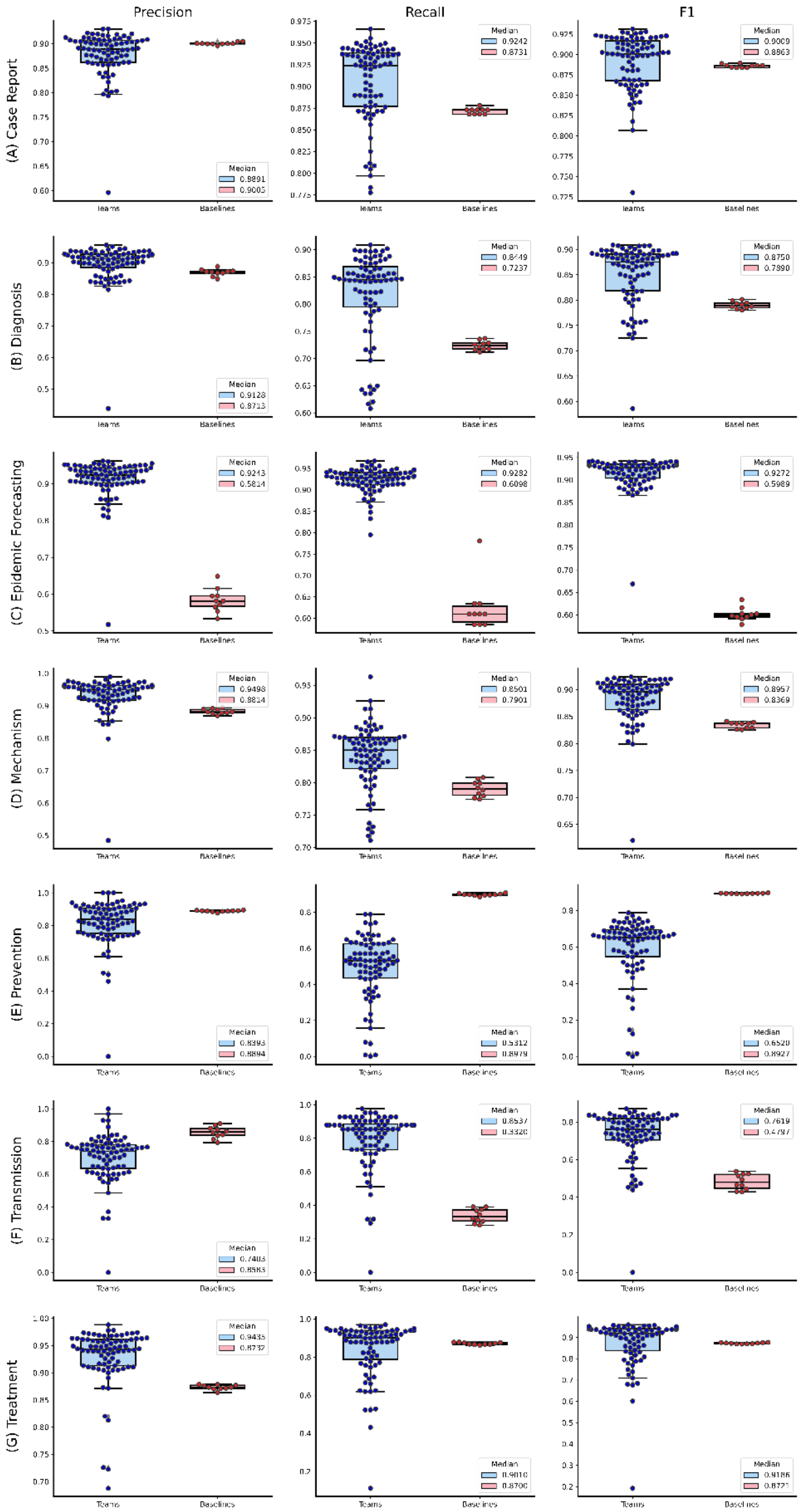

Figure 3. The distributions of team submission and baseline F1-scores for individual topics. Median F1-scores are shown in the legend.

Table 6 provides the top 5 team submission performance ranked by each of the F1-scores. The best score is 6.8%, 4.1%, and 4.1% higher than the corresponding team average score for macro F1-score, micro F1-score, and instance-based F1-score, respectively. Four teams (Bioformer, DonutNLP, DUT914, and polyu_cbsnlp) consistently achieved top-ranked performance in the three rankings. As mentioned above, the Bioformer and DUT914 teams proposed innovative methods which are beyond the default transfer learning approaches. In contrast, DonutNLP and polyu_cbsnlp used an ensemble of transformer approaches which also improve the performance. This is consistent with observations from previous challenge tasks [11, 24].

## Discussion and conclusions

This overview paper summarizes the BioCreative LitCovid track in terms of data collection and team participation. It provides a manually curated dataset of over 33,000 biomedical scientific articles. This is one of the largest datasets for multi-label classification for biomedical scientific literature, to our knowledge. Overall, 19 teams submitted 80 testing set predictions and ~75% of the submissions had better performance than the baseline approach. Given the scale of the dataset and the level of participation and team results, we conclude that the LitCovid track of BioCreative VII ran successfully and is expected to make significant contributions to innovative biomedical text mining methods.

One possible direction to explore is the efficiency of transformers in real-world applications. As described above, over 80% of the teams used the transformers; the top 5 team submissions also show superior performance using the transformer approach. However, it has a trade-off on the efficiency side. Existing studies show that transformers are significantly slower than other deep learning approaches using word and sentence embeddings, e.g., up to 80 times slower for biomedical sentence retrieval [84]. This is more challenging under the setting of multi-label classification (may require more than one transformer model) on COVID-19 literature (~10,000 articles per month). The Bioformer team showed one candidate approach, which only uses one third of the parameters used by the original transformer architecture and achieves similar performance. We expect more innovative transformer approaches will be developed to improve the efficiency.

Another possible direction is to quantify the usability of systems by incorporating them into the curation workflow. The systems are ultimately used to facilitate data curation – it is thus important to evaluate its usability in the curation workflow, e.g., what is the accuracy of systems for new articles and how much manual curation effort can be reduced by deploying the systems? We have conducted a preliminary analysis on the generalization capability and efficiency of the systems in the LitCovid production environment [85], and we encourage more studies to perform usability evaluation and accountability of systems in the curation workflow [86, 87].

A further possible direction is the development of datasets for biomedical multi-label classification tasks. As summarized above, while multi-label classification is frequently used in biomedical literature, limited datasets are available for method development. This seems the major bottleneck for innovative biomedical text mining methods. We expect a community effort for dataset construction and a combination of automatic and manual curation approaches would address this issue. Also, given the scale of the BioCreative LitCovid dataset, it would be interesting to explore whether it can support transfer learning to other biomedical multi-label classification tasks. We encourage further development of biomedical text mining methods using the BioCreative LitCovid dataset.

## Acknowledgment

This research is supported by the NIH Intramural Research Program, National Library of Medicine.